\documentclass[10pt]{article}

\usepackage{amsmath}

\usepackage{array}

\usepackage{appendix}

\usepackage{tocloft}                   

\usepackage{graphicx}

\usepackage{amsfonts}

\usepackage{amssymb}

\usepackage{mathrsfs}

\usepackage{yfonts}

\usepackage{euscript}

\usepackage{centernot}                 

\usepackage{ifsym}                     

\usepackage{upgreek}

\usepackage{mathtools}

\usepackage{color}

\usepackage{slantsc}
\usepackage{calligra}

\usepackage{bbold}          

\usepackage[T1]{fontenc}

\usepackage{epsf}

\usepackage{latexsym}

\usepackage{tipa}

\usepackage{makeidx}

\makeindex

\def\b{\begin{equation}}

\def\e{\begin{equation}}

\def\be{\begin{equation}}              

\def\ee{\end{equation}}

\def\beq{\begin{equation}}

\def\eeq{\end{equation}}

\def\bea{\begin{eqnarray}}

\def\eea{\end{eqnarray}}

\def\m{\mbox{ }}

\def\mma {\m , \m \m }

\def\!{\hspace{-1.6667em}}

\def\c{\cite}

\def\n{\noindent}

\def\u{\underline}

\def\w{\widetilde}

\def\s{\stackrel}

\def\bGamma{\mbox{\boldmath$\Gamma$}}               

\def\bA{\mbox{\bf A}}

\def\bQ{\mbox{\bf Q}}

\def\bT{\mbox{\bf T}}

\def\bg{\mbox{\bf g}}

\def\bsigma{\mbox{\boldmath$\sigma$}}                   %

\def\bGamma{\mbox{\boldmath$\Gamma$}}

\def\sumi2{\sum\mbox{}_{\mbox{}_{\mbox{\scriptsize $i$=1}}}^2}

\def\sumi3{\sum\mbox{}_{\mbox{}_{\mbox{\scriptsize $i$=1}}}^3}

\def\sumABcycles3{\sum\mbox{}_{\mbox{}_{\mbox{\scriptsize cycles $A,B$=1}}}^{3}}

\def\sumCDcycles3{\sum\mbox{}_{\mbox{}_{\mbox{\scriptsize cycles $C,D$=1}}}^{3}}

\def\sumj3{\sum\mbox{}_{\mbox{}_{\mbox{\scriptsize $j$=1}}}^3}

\def\sumk3{\sum\mbox{}_{\mbox{}_{\mbox{\scriptsize $k$=1}}}^3}

\def\prodiA1{\prod\mbox{}_{\mbox{}_{\mbox{\scriptsize $i$=1}}}^{A - 1}}

\def\bigtimes{\mbox{\Large $\times$}}

\def\d{\textrm{d}}                                                  

\def\es{\m = \m}

\def\:={\m := \m}

\def\=:{\m =: \m}

\def\FrC{\mbox{$\mathfrak{C}$}}                                
                                                       
\def\FrS{\mbox{\Large $\mathfrak{s}$}}                         
\def\FrM{\mbox{$\mathfrak{M}$}}                                
\def\lFrg{\mbox{\Large$\mathfrak{g}$}}                         

\def\Hilb{\mbox{{\boldmath$\mathfrak{H}$}ilb}}                 
\def\FrQ{\mbox{\Large $\mathfrak{q}$}}                               
\def\Phase{\mbox{{\boldmath$\mathfrak{P}$}hase}}                     

\def\bFrR{\mbox{\boldmath$\mathfrak{R}$}}                            
\def\Rig-Phase{\bFrR\mbox{ig-}\Phase}                                
                                                                													   
\def\FrP{\mbox{\Large $\mathfrak{p}$}}                                 
\def\FrR{\mbox{\boldmath$\mathfrak{R}$}}                             

\def\bFrR{\mbox{\boldmath$\mathfrak{R}$}}                            

\def\bFrR{\mbox{\boldmath$\mathfrak{R}$}}                            

\def\1mat{\u{\u{1}}}                                                 

\def\Positive-Modespace{\mbox{{\boldmath$\mathfrak{M}$}odespace$^+$}}

\def\POSITIVE-MODESPACE{\mbox{{\boldmath$\mathfrak{M}$}ODESPACE$^+$}}

\def\Kin-Hilb{\mbox{{\boldmath$\mathfrak{K}$}in-\Hilb}}                     

\def\Mid-Hilb{\mbox{{\boldmath$\mathfrak{M}$}id-\Hilb}}                     

\def\Dyn-Hilb{\mbox{{\boldmath$\mathfrak{D}$}yn-\Hilb}}                     

\def\5Star{\mbox{\Large$\star$}}              

\begin{document}

\begin{titlepage}

\begin{center}

{\bf\Large Shape Theories. I. Their Diversity is Killing-Based and thus Nongeneric}

\vspace{0.1in}
\m

{\bf Edward Anderson}$^*$  

\m 

\end{center}

\begin{abstract}

\n Kendall's Shape Theory covers shapes formed by $N$ points in $\mathbb{R}^d$ upon quotienting out the similarity transformations. 
This theory is based on the geometry and topology of the corresponding configuration space: shape space. 
Kendall studied this to build a widely useful Shape Statistics thereupon.
The corresponding Shape-and-Scale Theory -- quotienting out the Euclidean transformations -- is useful in Classical Dynamics and Molecular Physics, 
as well as for the relational `Leibnizian' side of the Absolute versus Relational Motion Debate.  
Kendall's shape spaces moreover recur withing this `Leibnizian' Shape-and-Scale Theory.  
There has recently been a large expansion in diversity of Kendall-type Shape(-and-Scale) Theories.
The current article outlines this variety, and furthermore roots it in solving the poset of generalized Killing equations.  
This moreover also places a first great bound on how many more Shape(-and-Scale) Theories there can be.
For it is nongeneric for geometrically-equipped manifolds -- replacements for Kendall's $\mathbb{R}^d$ carrier space (absolute space to physicists) -- 
to possess any generalized Killing vectors.  
Article II places a second great bound, now at the topological level and in terms of which Shape(-and-Scale) Theories are technically tractable.  
Finally Article III explains how the diversity of Shape(-and-Scale) Theories -- from varying which carrier space and quotiented-out geometrical automorphism group are in use -- 
constitutes a theory of Comparative Background Independence: a topic of fundamental interest in Dynamics, Gravitation and Theoretical Physics more generally.  
Article I and II's great bounds moreover have significant consequences for Comparative Background Independence.  

\end{abstract}

\n Mathematics keywords: Geometrical automorphism groups, generalized Killing equations, genericity, quotient spaces. 
Used for Shape Theory (underlying Shape Statistics and the below).

\m 

\n Physics keywords: Configuration spaces, Relational Mechanics, Background Independence, $N$-Body Problem, Minkowski Spacetime.

\m 

\n PACS: 04.20.Cv, 02.40.-k 

\m

\n $^*$ dr.e.anderson.maths.physics *at* protonmail.com 

\section{Introduction}

\subsection{Kendall's Shape Theory and its Leibnizian scaled counterpart}

General Kendall-type Shape(-and-Scale) Theory \cite{Kendall84, Kendall89, Small, Sparr, Kendall, MP03, MP05, FileR, Bhatta, DM16, PE16, ABook, I-II-III, IV-2, Minimal-N} 
is the culmination of the following four layers of structure. 

\m 

\n{\bf Structure I} {\it Carrier space} 
\be
\FrC^d              \m ,
\ee 
is an at least incipient model for the structure of space, alias {\it absolute space} in the context of physical modelling.   
While Geometry was originally conceived of as occurring in physical space or objects embedded therein (parchments, the surface of the Earth...), 
it was however subsequently reconceived as occurring in abstract space. 
We thus say `carrier space' rather than `absolute space' in the purely geometrical context. 
Carrier space can moreover also be interpreted as a {\it sample space} in the context of Probability-and-Statistics, of {\it location data}.  

\m 

\n Casting a fixed $d$-dimensional manifold in the role of carrier space, 
\be 
\FrC^d = \FrM^d    \m ,
\ee
is quite a general possibility (and one that the current Series of Articles resides entirely within).  
 
\m 
 
\n For Kendall's Shape Theory itself \cite{Kendall84, Kendall89, Kendall}, moreover, carrier space is just 
\be 
\FrC^d = \mathbb{R}^d    \m :  
\ee 
the `most obvious' case: all flat and bereft of topological nontriviality.  

\end{titlepage}

\n{\bf Structure II}  Both the Geometry and Probability-and-Statistics contexts involve studying {\sl multiple points on} carrier space. 
In some physical applications, moreover, the corresponding points on absolute space furthermore model material particles (classical, and taken to be of negligible extent).
To cover both of these situations at once in our exposition, we use a points-or-particles portmanteau concept. 
So in all three settings -- Geometry, Probability-and-Statistics, and Physics -- one is to study $N$-point-or-particle {\sl constellations}: 
a type of {\it configuration} \cite{Lanczos, Arnol'd}
\be 
\bQ = \u{Q}^I  \m ,  
\ee 
where the underline denotes carrier space vector and the point-or-particle label $I$ runs from 1 to $N$. 

\m 

\n{\it Configuration space} \cite{Lanczos, Arnol'd, AConfig}
\be 
\FrQ(\FrS)
\ee 
is then the abstract space formed from the totality of possible values that a given system $\FrS$'s configurations $\bQ$ can take.  

\m 

\n{\it Constellation spaces} 
\be 
\FrQ(\FrC^d, N)
\ee 
are the configuration spaces of constellations, on some fixed carrier space $\FrC^d$ and for some fixed count $N$ of points-or-particles. 

\m 

\n Constellation spaces are moreover particularly simple examples of configuration spaces, being just {\it finite product spaces} 
\be 
\FrQ(\FrC^d, N)  \es  \bigtimes_{i = 1}^N \FrC^d   \es   \bigtimes_{i = 1}^N \FrM^d          \m .     
\ee 
\n For Kendall's Shape Theory itself, these further simplify as follows. 
\be 
\FrQ(d, N)  :=   \FrQ(\mathbb{R}^d, N) 
            \es  \bigtimes_{I = 1}^N\mathbb{R}^d   
			\es  \mathbb{R}^{N \, d}                             \m . 
\ee 
\n{\bf Structure III} We furthermore consider some geometrically-significant group of automorphisms of $\FrC^d$,  
\be
\lFrg \:= Aut(\FrC^d, \bsigma) \es Aut(\FrM^d, \bsigma)          \m ,
\ee 
where $\bsigma$ is some level of geometrical structure that $\FrC^d = \FrM^d$ has been equipped with that is to be irrelevant to the modelling in question. 

\m 

\n{\bf Example 1} For $\sigma = \overline{\bg}$ the similarity structure -- i.e.\ metric structure up to constant rescalings -- on carrier space $\FrC^d = \FrM^d$, 
\be 
Aut(\FrM^d, \overline{\bg}) = Sim(\FrM^d)                         \m : 
\ee 
the {\it similarity group} of $\FrM^d$. 

\m 

\n In Kendall's case, $\FrC^d = \mathbb{R}^d$, this is the {\it similarity group of flat space},\footnote{More specifically, 
this is the continuous part thereof, meaning in this context that reflections have for now been omitted; see Sec 3.3.}
\be 
Sim(d) \es     Tr(d)     \rtimes (Rot(d) \times Dil) 
       \es  \mathbb{R}^d \rtimes (SO(d)  \times \mathbb{R}_+)     \m ,  
\ee 
of translations $Tr(d)$, rotations $Rot(d)$ and dilations $Dil$, where $\times$ and $\rtimes$ denote direct and semidirect product of groups \cite{Cohn} respectively.  
This is the case standardly considered in Kendall's Shape Statistics  
It turns out moreover to be a useful intermediary and/or structure within Example 1's Mechanics context as well, as explained at the end of this subsection.

\m 

\n{\bf Example 0}   For $\sigma = \bg$ the metric structure on carrier space $\FrC^d = \FrM^d$, 
\be 
Aut(\FrM^d, \bg) = Isom(\FrM^d)                                   \m : 
\ee 
the {\it isometry group} of $\FrM^d$. 

\m

\n In the case of $\FrC^d = \mathbb{R}^d$, this is the {\it Euclidean group}   
\be 
Eucl(d)  \es      Tr(d)    \rtimes Rot(d)    
         \es  \mathbb{R}^d \rtimes  SO(d) 
\ee 
of translations and rotations.
By now not involving the dilations, this is the {\it scaled counterpart}     of the similarity group. 
Conversely the similarity group          is the {\it pure-shape counterpart} of the Euclidean  group.
Finally, we say that, together, the similarity and Euclidean groups of flat space form a {\it shape(-and-scale) pair}.   

\m 

\n{\bf Structure IV} {\it Relational Theory} is then the study of some constellation space as quotiented by some geometrically-significant automorphism group of $\FrC^d$. 
This makes sense because $Aut(\FrC^d, \bsigma)$'s action on $\FrC^d$ readily uplifts to an action on $\FrQ(\FrC^d, N)$ by the latter's product space structure.  

\m 

\n If the geometrical automorphism group in question includes global scale (in a sense made more precise in Sec \ref{O-C}), 
then we say that the corresponding Relational Theory is a fortiori a {\it Shape Theory}. 

\m

\n If not, then we say that the corresponding Relational Theory is a fortiori a {\it Shape-and-Scale Theory}. 

\m 

\n `Relational Theory' is thus a portmanteau of Shape Theory and Shape-and-Scale Theory, also sometimes called {\it Shape(-and-Scale) Theory}. 

\m  

\n We refer to the corresponding configuration spaces as `shape space', `shape-and-scale space', and the portmanteau `shape(-and-scale)-space' alias `relational space'. 
These are denoted by $\FrS$, $\FrS\FrS$, and $\FrR$ respectively.  

\m 

\n{\bf Example 1} For carrier space $\mathbb{R}^d$, 
quotienting out constellation space by the similarity group $Sim(d)$ gives {\it Kendall's Similarity Shape Theory} \cite{Kendall84}.  
The corresponding shape spaces are \cite{Kendall84, Kendall}
\be 
\FrS(d, N)  \:=   \frac{  \FrQ(d, N)  }{  Sim(d)  } 
            \es   \frac{  \bigtimes_{i = 1}^N \mathbb{R}^d  }{  Tr(d) \rtimes (Rot(d) \times Dil)  } 
            \es   \frac{  \mathbb{R}^{N \, d}  }{  \mathbb{R}^d \rtimes (SO(d) \times \mathbb{R}_+)  }  
            \es   \frac{  \mathbb{R}^{n \, d}  }{  SO(d) \times \mathbb{R}_+  } 
            \es   \frac{  \mathbb{S}^{n \, d - 1}  }{  SO(d)  } 		                                          \m , 
\ee
where 
\be 
n : = N - 1
\ee 
is the number of independent relative separations.  

\m 

\n{\bf Example 0} For carrier space $\mathbb{R}^d$, quotienting out the constellation space by the Euclidean group $Eucl(d)$ gives 
{\it `Leibnizian'} alias {\it Metric Shape-and-Scale Theory} \cite{Smale70, BB82, LR95-LR97, M02-M05, FileR, Minimal-N}.
The corresponding shape-and-scale spaces are 
\be 
\FrS\FrS(d, N)  \:=   \frac{  \FrQ(d, N)}{Isom(\mathbb{R}^d)}
                \es   \frac{  \bigtimes_{i = 1}^N \mathbb{R}^d}  {  Eucl(d)  } 
                \es   \frac{  \mathbb{R}^{N \, d}  }{  \mathbb{R}^d \rtimes SO(d)  }  
                \es   \frac{  \mathbb{R}^{n \, d}  }{  SO(d)  }                                                       \m .
\ee 
This is the most usually considered setting for the relational horn of the venerable Absolute versus Relational (Motion) Debate \cite{Newton, L, M, BB82, DoD-Buckets, ABook}, 
often contemporarily also referred to as `seeking Background Independence' \cite{A64-A67, Battelle, DeWitt67, K92, I93, Giu06, APoT, ABook}, 
or, in the event of this being difficult to implement, as `the Problem of Time'.  

\m 

\n Examples 1 and 0 additionally constitute a first Shape(-and-Scale) Theory pair. 
In studying generalized Kendall-type Relational Theory, such pairs are initially a common occurrence, but one subsequently finds out that they are exceedingly rare. 
Because of this, we also coin a term for those Relational Theories which have no such pair: {\it singleton theories}.  

\m 

\n Some subcases of Examples 1 and 0 moreover simplify topologically and geometrically as follows.   

\m 

\n{\bf Example 1.1} For Similarity Shape Theory in 1-$d$, straightforwardly 
\be 
\FrS(N, 1) = \mathbb{S}^{n - 1}  \m \mbox{(spheres)}  \m 
\ee 
both topologically and metrically.  
This moreover generalizes to 
\be 
\FrP(N, d) = \mathbb{S}^{n \, d - 1}
\ee 
-- Kendall's preshape sphere -- being the result of quotienting out translations and dilations -- but not rotations -- from constellation space.  
This explains the last step in (eq) via  
\be 
\FrS(N, d) \es \frac{\FrP(N, d)}{Rot(d)}   \m . 
\ee 
The last expression in (eq) moreover does not in general simplify. 

\m 

\n{\bf Example 1. 2} For Similarity Shape Theory in 2-$d$, however, \cite{Smale70}
\be 
\FrS(N, 2) \es \frac{\mathbb{S}^{n \, d - 1}}{SO(2)} \es  \frac{\mathbb{S}^{n \, d - 1}}{\mathbb{S}^1} \es \mathbb{CP}^{n - 1} \m \mbox{(complex-projective spaces)} \m . 
\ee
This is by the generalized Hopf map (or alternative workings, as reviewed in \cite{Quad-I}).  
In its realization as a shape space, this is additionally equipped with the standard Fubini--Study metric \cite{Kendall84}. 
For $N = 3$, this furthermore reduces to the {\it shape sphere} of triangles, by the topological and geometrical coincidence 
\be
\mathbb{CP}^1 = \mathbb{S}^2 \m .  
\ee 
For $N = 3$ in 3-$d$, the working shares many similarities but the outcome is a shape hemisphere with edge, as reviewed in \cite{A-Monopoles}. 

\m

\n{\bf Example 0.1)} For Euclidean Shape-and-Scale Theory in 1-$d$,  the corresponding {\it Leibnizian shape-and-scale space} is 
\be
\FrS\FrS(N, 1) = \mathbb{R}^{n}  \m .
\ee 
These are the topological and metric cones over the corresponding shape spaces, a result which holds in general for Euclidean shape-and-scale spaces \cite{FileR}.

\subsection{Outline of the rest of this Article and Series}\label{Cone}

\n We next survey the large recent increase in diversity among Shape(-and-Scale) Theories, alias Relational Theory.
We start by varying the geometrically meaningful automorphism group quotiented out in Sec 2.1. 
We next vary the carrier space itself in Sec 2.2. 
We finally briefly contemplate quotienting out discrete transformations as well in Sec 2.3, and less structured rubber shapes in Sec 2.4.

\m 

\n We next reveal further foundations for Relational Theory in Sec 3, in the form of the poset \cite{PE-1} of generalized Killing equations \cite{Yano}. 
On the one hand, it is by solving these equations that current and subsequent Relational Theories can be found systematically. 
On the other hand, it is well-known that these equations generically admit no nontrivial solutions on the geometrically-equipped differentiable manifolds. 
For these solutions are generalized Killing vectors, and such are only supported by geometrically-equipped differentiable manifolds that possess generalized symmetries, 
of which the generic such manifold possesses none.  
There are thus generically {\sl no} nontrivial Relational Theories 
(trivial ones being those whose configuration spaces are mere constellation space products rather than nontrivial quotients thereof). 

\m 

\n Within Article I's nongeneric case in which there {\sl are} nontrivial Relational Theories, 
Article II points to almost all of these having not manifolds but {\sl stratified manifolds} \c{W-T} for their relational spaces.  
This is because quotienting by Lie groups in general kicks one out of the category of manifolds into that of stratified manifolds, 
of which quotienting constellation spaces by geometrically-significant automorphism groups to form relational spaces is a subcase.   
We point moreover to there being three classes of technical tractability for such stratified manifolds, giving representatives of each among hitherto studied Relational Theories. 
We furthermore underlie this by topological-level protection theories, giving compactness criteria that guarantee the most technically tractable class.  
This places a second great bound -- now based on technical tractability -- to Article I's great bound at the level of generic non-existence.  

\m 

\n We can moreover place a Mechanics on each of these examples \cite{FORD, Cones, FileR, AMech, ASphe, ATorus, Forth}\footnote{These works also show that reducing the indirectly formulated \cite{BB82, B03} 
in the globally tractable 1- and 2-$d$ cases returns Kendall's shape spaces and the cones thereover as the corresponding reduced configuration spaces.  
Similar considerations apply, albeit now only locally, in $\geq 3$-$d$.
\cite{LR95-LR97} and \cite{M02-M05} furthermore consider the role of (scaled) shape spaces context of subsystem models, for such applications as Celestial Mechanics and Molecular Physics.}
%
Entertaining these extends consideration of the relational horn of the Absolute versus Relational Debate, 
now giving a {\sl Comparative} theory of Background Independence \cite{AMech, ABook, A-Generic, A-CBI}.  
Article III then considers Article I and II's great bounding results in the light of the Comparative Background Independence program, 
which generalizes the Absolute versus Relative Debate to arbitrary carrier spaces and geometrically-significant automorphism groups acting thereupon.
This also serves as an opportunity to introduce some fundamental Theoretical Physics notions of Background Independence to shape theorists (mostly Statisticians and Applied Mathematicians), 
given that their subject actually provided the first keys to formulating an actual full-blown theory of Background Independence... 
[This unlocking has moreover rolled on \cite{ABook}, across phase space and spacetime, and into the quantum world.]  
For readers who are, conversely, theoretical physicists seeking to understand Shape Theory, consult also e.g. \cite{FORD, AKendall, A-Generic} 
and the Reviews in this Introduction's opening concatenation of references. 

\vspace{10in}

\section{Large recent increase in diversity of (Scale-and-)Shape Theories} 

\subsection{Quotienting out other continuous automorphism groups $\lFrg$}

\n There has recently been a large increase in diversity of (Scale-and-)Shape Theories; again, the first motivation bringing this about was Shape Statistics \c{Kendall87, MP03, MP05}. 
Subsequently Mechanics \cite{AMech}, Background Independence \cite{AMech, ABook} and Differential Geometry \cite{Shape-Derivatives}.  
There are various `axes of generalization' as embodied by the below subsections; one can moreover apply whichever combination of these generalizations.  

\m 

\n{\bf Example 3} The affine group in \cite{Sparr, MP03} with subsequent key work in \cite{GT09} and reviews in \cite{Bhatta, PE16, Aff}.  
For $\sigma = \bA$ the affine structure, 
\be 
Aut(\FrC^d, \bA) = Aff(\FrC^d)                                                      \m : 
\ee 
the {\it affine group} of $\FrC^d$. 
In the case of $\FrC^d = \mathbb{R}^d$, this is the affine group 
\be 
Aff(d) = Tr(d) \rtimes GL(d, \mathbb{R})                                            \m , 
\ee 
for $GL(d, \mathbb{R})$ the {\it general-linear group} of dilations, rotations, shears $Sh(d)$ and Procrustes stretches $Pr(d)$ \cite{Coxeter}. 
Quotienting out by the affine group $Aff(d)$ then gives {\it Affine Shape Theory}, whose shape space is  
\be 
\FrS(d, N; Aff)     \:=   \frac{\FrQ(d, N)}{Aff(d)} 
                    \es   \frac{\FrQ(d, N)}{Tr(d) \rtimes GL(d, \mathbb{R})} 
                    \es   \frac{\bigtimes_{i = 1}^N \mathbb{R}^d}{\mathbb{R}^d \rtimes GL(d, \mathbb{R})}    
			        \es   \frac{\mathbb{R}^{N \, d}}{GL(d, \mathbb{R})} 
                    \es	  \frac{\mathbb{S}^{N \, d - 1}}{SL(d, \mathbb{R})}			\m .
\ee 
This is the case corresponding to Image Analysis from the idealized perspective of an infinitely distant observer \cite{Sparr, MP03, PE16}.  

\m 

\n Affine Shape Theory moreover has a scaled partner, {\it Equi-top-form Shape-and-Scale Theory}.\footnote{Equiareal Geometry \c{Coxeter} 
-- the 2-$d$ case -- is the most familiar Equi-top-form Geometry.} 
%
This case's corresponding automorphism group is 
\be 
Equi(d) = Tr(d) \rtimes SL(d, \mathbb{R})                                            \m , 
\ee 
for $SL(d, \mathbb{R})$ the {\it special-linear group} [scale-free version of $GL(d, \mathbb{R})$].  
The corresponding shape-and-scale space is 
\be 
\FrS\FrS(d, N; Equi)     \:=   \frac{\FrQ(d, N)}{Equi(d)} 
                         \es   \frac{\FrQ(d, N)}{Tr(d) \rtimes SL(d, \mathbb{R})} 
                         \es   \frac{\bigtimes_{i = 1}^N \mathbb{R}^d}{\mathbb{R}^d \rtimes SL(d, \mathbb{R})}    
			             \es   \frac{\mathbb{R}^{N \, d}}{SL(d, \mathbb{R})}                              			\m .
\ee 
$\mathbb{R}^d$ only supports affine and equiareal groups that are distinct from the similarity and Euclidean groups respectively if $d \geq 2$.  

\m 

\n{\bf Example 4} Projective Shape Theory \cite{MP05, KKH16} and reviews in \cite{Bhatta, PE16, Forth}. 
Details are postponed until the next section for technical reasons. 

\m 

\n{\bf Example 5} Conformal Shape Theory   
For $\sigma = \w{\bg}$ the conformal structure, 
\be 
Aut(\FrC^d, \w{\bg}) = Conf(\FrC^d)                             \m : 
\ee 
the {\it conformal group} of $\FrC^d$. 
In the case of $\FrC^d = \mathbb{R}^d$ for $d \geq 3$, this is the conformal group 
\be 
Conf(d) = SO(d, 1)
\ee
of $Tr(d)$, $Rot(d)$, $Dil$, and special conformal transformations $K(d)$ \cite{O17}. 
Quotienting out by the conformal group $Conf(d)$ gives {\it Conformal Shape Theory} \cite{AMech, ABook}, whose shape space is  
\be 
\FrS(d, N; Conf)  \:=   \frac{\FrQ(d, N)}{Conf(\mathbb{R}^d)}  
                  \es   \frac{\bigtimes_{i = 1}^N \mathbb{R}^d}{SO(d, 1)}                                                \m .
\ee 
\n{\bf Remark 1} While the affine case has a scaled partner, the projective and conformal cases are singleton theories, 
since their Lie brackets structure precludes quotienting out dilations by themselves \c{A-Brackets}. 
`Passing to centre of mass frame' is also no longer possible for these, on similar grounds.

\subsection{Other carrier spaces $\FrC^d$}\label{O-C}

\n{\bf Example 6} The circle \cite{Roach, JM00, ACirc}
\be 
\FrC^d = \mathbb{S}^1                    \m .  
\ee 
This is moreover not only the first sphere                 
but also the                   first torus                  
and the                        first real-projective space: 
\be 
\mathbb{T}^1 = \mathbb{S}^1 = \mathbb{RP}^1                                                                       \m .
\ee 
The corresponding constellation space is 
\be 
\FrQ(\mathbb{S}^1, N)  \:= \bigtimes_{I = 1}^N \mathbb{S}^1 \es \mathbb{T}^N \m :
\ee
a torus. 

\m 
 
\n The circle moreover supports an isometry group, 
\be 
Isom(\mathbb{S}^1) = SO(2) 
                   = U(1) 
				   = \mathbb{S}^1                                                                                 \m , 
\ee 
where the last equality is at the level of manifolds. 
This consists of periodically-identified translations.  
\m 

\n We thus have a {\it Metric Shape-and-Scale Theory} on the circle, with shape-and-scale space  
\be 
\FrS\FrS(\mathbb{S}^1, N)  \:=   \frac{\FrQ(\mathbb{S}^1, N)}{Isom(\mathbb{S}^1)}
                       \es   \frac{\bigtimes_{i = 1}^N \mathbb{S}^1}{\mathbb{S}^1}  
                       \es   \bigtimes_{i = 1}^n \mathbb{S}^1
                       \es   \mathbb{T}^n	                         				                              \m : 
\ee 
another torus.  

\m 

\n In this case, moreover, the projective group and the conformal group coincide: 
\be 
Proj(\mathbb{S}^1) = SO(2, 1) = PSL(2, \mathbb{R}) =  Conf(\mathbb{S}^1)  \m .  
\ee 
This gives 
\be 
\FrS\FrS(\mathbb{S}^1, N; Proj)  \es  \FrS\FrS(\mathbb{S}^1, N; Conf) 
                                 \es  \frac{\FrQ(\mathbb{S}^1, N)}{PSL(2, \mathbb{R})} 
								 \es  \frac{\mathbb{T}^N}{PSL(2, \mathbb{R})}
\ee 

\n Isometries for all of the below are singleton theories: there are no corresponding distinct Metric Shape Theories on any of these carrier spaces.  

\m 

\n{\bf Example 7} The $d \geq 2$ tori \cite{JM00, ATorus} 
\be 
\FrC^d   =   \mathbb{T}^d  
        \es  \bigtimes_{I = 1}^N \mathbb{S}^1                                                                     \m .  
\ee
The corresponding constellation spaces are 
\be 
\FrQ(\mathbb{T}^d, N)  \:=  \bigtimes_{I = 1}^N \mathbb{T}^d  
                       \es  \bigtimes_{I = 1}^N \bigtimes_{a = 1}^d \mathbb{S}^1  
					   \es  \bigtimes_{I = 1}^{N \, d}              \mathbb{S}^1   \es \mathbb{T}^{N \, d}        \m :
\ee
a bigger torus. 

\m 

\n These tori moreover support the isometry groups 
\be 
Isom(\mathbb{T}^d)  \es  \bigtimes_{a = 1}^d U(1)                                                                    
                    \es  \bigtimes_{a = 1}^d \mathbb{S}^1 
					\es  \mathbb{T}^d                       \m . 
\ee
We thus have a {\it Metric Shape-and-Scale Theory}, with shape-and-scale spaces   
\be 
\FrS\FrS(\mathbb{T}^d, N)  \:=   \frac{\FrQ(\mathbb{T}^d, N)}{Isom(\mathbb{T}^d)}
                           \es   \frac{\bigtimes_{i = 1}^{N \, d} \mathbb{S}^1}{\bigtimes_{i = 1}^{d} \mathbb{S}^1}  
                           \es   \bigtimes_{i = 1}^{n \, d} \mathbb{S}^1
                           \es   \mathbb{T}^{n \, d}                    	                         				  \m .
\label{Cancel4}
\ee
\n{\bf Example 8} The $d \geq 2$ spheres \cite{Kendall87, FileR, ASphe, ASphe2} 
\be
\FrC^d   =   \mathbb{S}^d                                                                                         \m .
\ee
The corresponding constellation spaces are 
\be 
\FrQ(\mathbb{S}^d, N)  \:=  \bigtimes_{I = 1}^N \mathbb{S}^d                                                      \m .
\ee
These spheres support the isometry groups 
\be 
Isom(\mathbb{S}^d) = SO(d + 1)                                                                                   \m ,  
\ee
giving {\it Metric Shape-and-Scale Theory}, with shape-and-scale space  
\be 
\FrS\FrS(\FrS^d, N)  \:=  \frac{\FrQ(\mathbb{S}^d, N)}{Isom(\mathbb{S}^d)}
                 \es  \frac{\bigtimes_{i = 1}^N \mathbb{S}^d}{SO(d + 1)}                                         \m .
\ee
So e.g.\ the $d = 2$ case's $SO(3)$ admits a $SO(2) = U(1)$ subgroup for the merely axisymmetrically-identified configurations.

\m 

\n These spheres moreover support a distinct conformal group, 
\be 
Conf(\mathbb{S}^d) = SO(d + 1, 1)
\ee 
giving {\it Conformal Shape-and-Scale Theory}, with relational space  
\be 
\FrS\FrS(\mathbb{S}^d, N; Conf)  \:=  \frac{\FrQ(\mathbb{S}^d, N)}{Conf(\mathbb{S}^d)}
                                 \es  \frac{\bigtimes_{i = 1}^N \mathbb{S}^d}{SO(d + 1, 1)}                                     \m .
\ee
\n These spheres do not support a distinct affine group, but do support a distinct projective group, 
\be 
Proj(\mathbb{S}^d) = PSL(d + 1) \m . 
\ee 
Note that for $d \geq 2$, unlike for $d = 1$'s circle,  
\be 
SO(d + 1, 1) \m \not{{\hspace{-0.04in}\cong}} \m PSL(d + 1)  \m , 
\ee 
as is already clear by dimensional difference: 
\be 
(d + 2)(d + 1)/2 = d^2 - 1  \m \Rightarrow \m  (d + 2)(d - 1) = 0 
                            \m \Rightarrow \m  d = 1 \m \mbox{(or $-2$, unrealized)} \m .
\ee 
The corresponding {\it Projective Shape-and-Scale Theory} has relational space 
\be 
\FrS\FrS(\FrS^d, N; Proj)  \:=  \frac{\FrQ(\mathbb{S}^d, N)}{Proj(\mathbb{S}^d)}
                           \es  \frac{\bigtimes_{i = 1}^N \mathbb{S}^d}{PSL(d + 1)}                                     \m .
\ee
\n{\bf Example 9} The $d \geq 2$ real projective spaces \cite{MP05, Bhatta, PE16, KKH16, PE-2-3, Forth} 
\be 
\FrC^d =  \mathbb{RP}^d    \m .  
\ee
The corresponding constellation spaces are 
\be 
\FrQ(\mathbb{RP}^d, N)  \:=  \bigtimes_{I = 1}^N \mathbb{RP}^d                                                   \m .
\ee
\n These real-projective spaces moreover support the isometry groups 
\be 
Isom(\mathbb{RP}^d)  = SO(d + 1) \m . 
\ee
We thus have a {\it Metric Shape-and-Scale Theory}, with shape-and-scale spaces   
\be 
\FrS\FrS(\mathbb{RP}^d, N)  \:=   \frac{\FrQ(\mathbb{T}^d, N)}{Isom(\mathbb{T}^d)}
                       \es   \frac{\bigtimes_{i = 1}^{N \, d} \mathbb{S}^1}{\bigtimes_{i = 1}^{d} \mathbb{S}^1}  
                       \es   \bigtimes_{i = 1}^{n \, d} \mathbb{S}^1
                       \es   \mathbb{T}^{n \, d}                    	                         				  \m .
\label{Cancel}
\ee
\n $\mathbb{RP}^d$ additionally supports a projective transformation group 
\be 
Proj(\mathbb{RP}^{d - 1}) = PGL(d, \mathbb{R}) \m .
\ee 
It is actually this that enters Image Analysis from the general and directly-realized perspective of a finitely-placed observer \cite{MP05, Bhatta, PE16}.  
The corresponding relational space is then \cite{MP05}
\be 
\FrS\FrS(\mathbb{RP}^{d - 1}, N; Proj)  \es  \frac{\bigtimes_{I = 1}^N \mathbb{RP}^{d - 1}}{PGL(d, \mathbb{R})}   \m .
\ee
\n{\bf Example 10} The $\d \geq 2$ hyperbolic spaces 
\be 
\FrC^d   =   \mathbb{H}^d                                         \m .  
\ee
The corresponding constellation spaces are 
\be 
\FrQ(\mathbb{H}^d, N)  \:=  \bigtimes_{I = 1}^N \mathbb{H}^d      \m .  
\ee
\n These support the isometry groups 
\be 
Isom(\mathbb{H}^d)  \es  SO(d, 1)                                \m , 
\ee
giving a {\it Metric Shape-and-Scale Theory}, with shape-and-scale spaces   
\be 
\FrS\FrS(\mathbb{H}^d, N)  \:=   \frac{\FrQ(\mathbb{H}^d, N)}{Isom(\mathbb{H}^d)}
                       \es   \frac{\bigtimes_{i = 1}^{N} \mathbb{H}^d}{SO(d, 1)}                    				  \m .
\label{Cancel2}
\ee
\n These hyperbolic spaces moreover support a distinct conformal group \cite{Bengtsson}, 
\be 
Conf(\mathbb{H}^d) = SO(d + 1, 1)   \m , 
\ee 
giving {\it Conformal Shape-and-Scale Theory}, with relational space  
\be 
\FrS\FrS(\mathbb{H}^d, N; Conf)  \:=  \frac{\FrQ(\mathbb{H}^d, N)}{Conf(\mathbb{H}^d)}
                                 \es  \frac{\bigtimes_{i = 1}^N \mathbb{H}^d}{SO(d + 1, 1)}                                     \m .
\ee
\n{\bf Example 11} Supersymmetric carrier space                    
\be 
\FrC^d   =   \mathbb{R}^{(p|q)} \mma p + q = d > 1                                                                   \m :   
\ee
a Grassmann space. 

\m 

\n The corresponding constellation spaces are 
\be 
\FrQ(\mathbb{R}^{(p|q)}, N)  \:= \bigtimes_{I = 1}^N \mathbb{R}^{(p|q)}                                               \m .
\ee
These support geometrical automorphism supergroups, among which \cite{AMech} considers 
\be 
Super\mbox{-}Tr(1)  \m .  
\ee 
The corresponding {\it Metric Shape-and-Scale Theory} has shape-and-scale space   
\be 
\FrS\FrS(\mathbb{R}^{(1|1)}, N)  \:=  \frac{  \FrQ(\mathbb{R}^{(1|1)}, N)  }{  Super\mbox{-}Tr(1)  }
                             \es  \frac{  \bigtimes_{I = 1}^N \mathbb{R}^{(1|1)}  }{  Super\mbox{-}Tr(1)  }             \m .
\label{Cancel3}
\ee
\n{\bf Example 12} Minkowski spacetime in the role of carrier space 
\be 
\FrC^D   =   \mathbb{M}^{D} = \mathbb{R}^{d, 1}  \mma  d \geq 1                                                         \m .  
\ee
The corresponding event constellation spaces are 
\be 
\FrQ(\mathbb{M}^D, N)  \:=  \bigtimes_{I = 1}^N \mathbb{R}^{d, 1} \es \mathbb{R}^{N \, d , \, N}                   \m .
\label{M-D}
\ee
$\mathbb{R}^{a,b}$ here denotes the indefinite flat space with $a$ + signs and $b$ $-$ signs.
These support geometrical automorphism supergroups, including the $(d, 1)$ in place of $d$ versions of Examples 0, 1, 2, 3, 4, 5-and-9 and 11.
Quotienting (\ref{M-D}) by each of these gives a distinct spacetime relational theory.

\subsection{Introducing a discrete quotient group}  

\c{FileR, PE16, I-II-III} include review of the effect of quotienting out by a reflection ($\mathbb{Z}_2\mbox{-ref}$ group).  

\m 

\n \cite{FileR, Quad-I, I-II-III} consider also indistinguishable labels ($S_N$ permutation group).  

\m 

\n \cite{A-Monopoles} extends this to partially distinguishable labels as well (subgroups of $S_N$).  

\m 

\n This subsection's modelling is however largely left for another occasion rather than developed in the current Series.

\subsection{Rubber shapes}

\n{\bf Example 12} Rubber shapes \cite{Top-Shapes} These have merely topological rather than metrical properties. 
This still distinguishes between different types of coincidence-and-collision, including the general shapes that are free from any such (but are now usually a single rubber shape). 
It moreover has three universal types for connected Hausdorff carrier spaces: 
$\mathbb{R}$, $\mathbb{S}^1$ and {\sl all such manifolds with dimension} $\geq$ 2 {\sl as a single model}. 
This distinction is because the order in which distinguishably labelled points-or-particles sit on a line is topologically encoded, 
and with some distinction between the $\mathbb{R}$ and $\mathbb{S}^1$ cases. 

\mbox{  }

\n This example is moreover sufficiently less structured to be qualitatively different from all other examples in Articles I and II. 
E.g.\ rubber relational spaces are not stratified manifolds but {\sl graphs}, 
and generalized Killing equations' differential-geometric level technology plays no part here in determining what automorphism groups are admissible.

\section{Foundations for Relational Theories }

\subsection{Poset of generalized Killing equations} 

\n Let $\FrM$ be a differentiable manifold.  
Consider an infinitesimal transformation 
\be 
\u{x} \longrightarrow \u{x}^{\prime} + \epsilon \, \u{\xi} \m 
\label{inf}
\ee 
thereupon.
For this to preserve an object, say a tensor, $\bT$, substituting (\ref{inf}) into $\bT$'s transformation law and equating first-order terms in $\epsilon$ gives 
\be 
\pounds_{\u{\xi}}\bT = 0 \m .  
\ee
$\pounds_{\u{\xi}}$ is here the Lie derivative \cite{Yano} with respect to $\u{\xi}$: a differential operator guaranteed by $\FrM$'s differential structure. 
Below we cast a fortiori geometrically significant objects in the role of $\bT$. 

\m 

\n{\bf Definition 1} For $\langle \, \FrM, \, \bsigma \, \rangle$ a manifold $\FrM$ equipped with a geometrically-significant level of structure $\bsigma$, 
the {\it generalized Killing equation (GKE)} is 
\be 
{\cal GK} \, \u{\xi} :=  \pounds_{\u{\xi}} \bsigma  =  0  \m . 
\ee 
This is moreover to be regarded as a PDE to solve for $\u{\xi}$: the {\it generalized Killing vectors (GKV)}.  
These form the algebra corresponding to the continuous part of the automorphism group $Aut(\FrM, \bsigma)$.    

\m 

\n{\bf Example A} The most familiar case is as follows. 
For  $\langle \, \FrM, \, \bg \, \rangle$ an (arbitrary-signature) Riemannian manifold, i.e.\ carrying a Riemannian metric structure $\bg$, 
\be 
{\cal K} \, \u{\xi} :=  \pounds_{\u{\xi}} \bg = 0  \m 
\label{LKE}
\ee   
is {\it Killing's equation}. 
Its solutions $\u{\xi}$ are {\it Killing vectors}.  
These are of significance in the study of  $\langle \, \FrM, \, \bg \, \rangle$ as they correspond to the {\sl symmetries}, 
more precisely to the {\sl isometries}, the totality of Killing vectors forming the {\it isometry group} 
\be 
Isom(\FrM, \, \bg )  \m .
\ee
\n{\bf Example B} For  $\left\langle \, \FrM, \, \overline{\bg} \, \right\rangle$ a manifold equipped with metric structure modulo constant rescalings, $\overline{\bg}$,
\be 
{\cal SK} \u{\xi}  :=  \pounds_{\u{\xi}} \overline{\bg} = 0 
\label{LSKE}
\ee 
is the {\it similarity Killing equation} \cite{MacCallum} (alias {\it homothetic Killing equation} in \cite{Yano}).  
Its solutions $\u{\xi}$ are {\it similarity Killing vectors}, consisting of isometries alongside 1 or 0 {\it proper similarities}, i.e.\ non-isometries.   
The totality of these form the {\it similarity group} 
\be 
Sim \left( \FrM, \, \overline{\bg} \right)                                             \m ;  
\ee 
for $\langle \, \FrM, \, \bg \, \rangle$ possessing no proper similarities, 
\be
Sim \left( \FrM, \, \overline{\bg} \right) = Isom \left( \FrM, \, \bg \right)  \m .  
\ee
\n{\bf Example C} For  $\left\langle \, \FrM, \, \widetilde{\bg} \, \right\rangle$ a manifold equipped with metric structure modulo local rescalings, $\w{\bg}$, 
\be 
{\cal CK} \, \u{\xi} :=  \pounds_{\u{\xi}} \widetilde{\bg} = 0 
\label{LCKE}
\ee 
is the {\it conformal Killing equation}.  
Its solutions $\u{\xi}$ are {\it conformal Killing vectors}, consisting of similarities and special conformal transformations. 
This is for $d \geq 3$; 
for $\mathbb{R}^2$ or $\mathbb{C}$', we get an infinity of analytic functions by the flat conformal Killing equation collapsing in 2-$d$ to the Cauchy--Riemann equations \cite{AMP}.
The totality of conformal Killing vectors form the {\it conformal group}
\be 
Conf( \FrM, \, \w{\bg} )  \m .  
\ee  
\n{\bf Example D} For  $\langle \, \FrM, \, {\bGamma} \, \rangle$ a manifold equipped with affine structure,  $\bGamma$
\be 
{\cal AK} \, \u{\xi} :=  \pounds_{\u{\xi}} \bGamma = 0 
\label{LAKE}
\ee 
is the {\it affine Killing equation}. 
Its solutions $\u{\xi}$ are {\it affine Killing vectors}, consisting of similarities alongside shears and squeezes alias Procrustes stretches \cite{Coxeter}.  
The totality of these form the {\it affine group} 
\be 
Aff( \FrM, \, \bGamma )            \m . 
\ee 
\n{\bf Example E} For  $\left\langle \, \FrM, \, \s{P}{\bGamma} \, \right\rangle$ a manifold equipped with projective structure,  

\n\be 
{\cal PK} \, \u{\xi}  := \pounds_{\u{\xi}} \s{P}{\bGamma} = 0 
\label{LPKE}
\ee 
is the {\it projective Killing equation}.    
Its solutions $\u{\xi}$ are {\it projective Killing vectors}, consisting of affine transformations alongside 
`special projective transformations'.    
The totality of these form the {\it projective group} 

\n\be 
Proj\left( \FrM, \, \s{P}{\bGamma} \right)           \m . 
\ee 
Having built up these examples, we entertain some discussion on nomenclature. 
Killing found the first such equation; its solutions form the isometry group of a geometry. 
{\it Generalized Killing equation (GKE)} is then a collective name for the five equations (\ref{LKE}, \ref{LSKE}, \ref{LCKE}, \ref{LAKE}, \ref{LPKE}) (and more \cite{Yano}), 
and  {\it generalized Killing vectors (GKV)} for the corresponding solutions.   
A more conceptually descriptive name is, moreover, {\it automorphism equation}, as explained in \cite{PE-1, PE-2-3, 9-Pillars}, 
whose solutions are now of course just termed {\it automorphism generators}. 

\m 

\n Given a carrier space, we go through the corresponding poset of GKEs to get the admissible geometrically significant automorphism groups. 
This provides a systematic way of arriving at the previous two sections' examples, 
as well as precluding various further possibilities not mentioned there from being distinct. 
The overall output is the posets of Fig \ref{Killing1}; 
considering projective and conformal groups separately as `top groups' \cite{A-Brackets} of geometrically-meaningful automorphisms in flat space, 
one is dealing with bounded lattices of geometrically significant subgroups. 
%
{\begin{figure}[ht]
\centering
\includegraphics[width=1.0\textwidth]{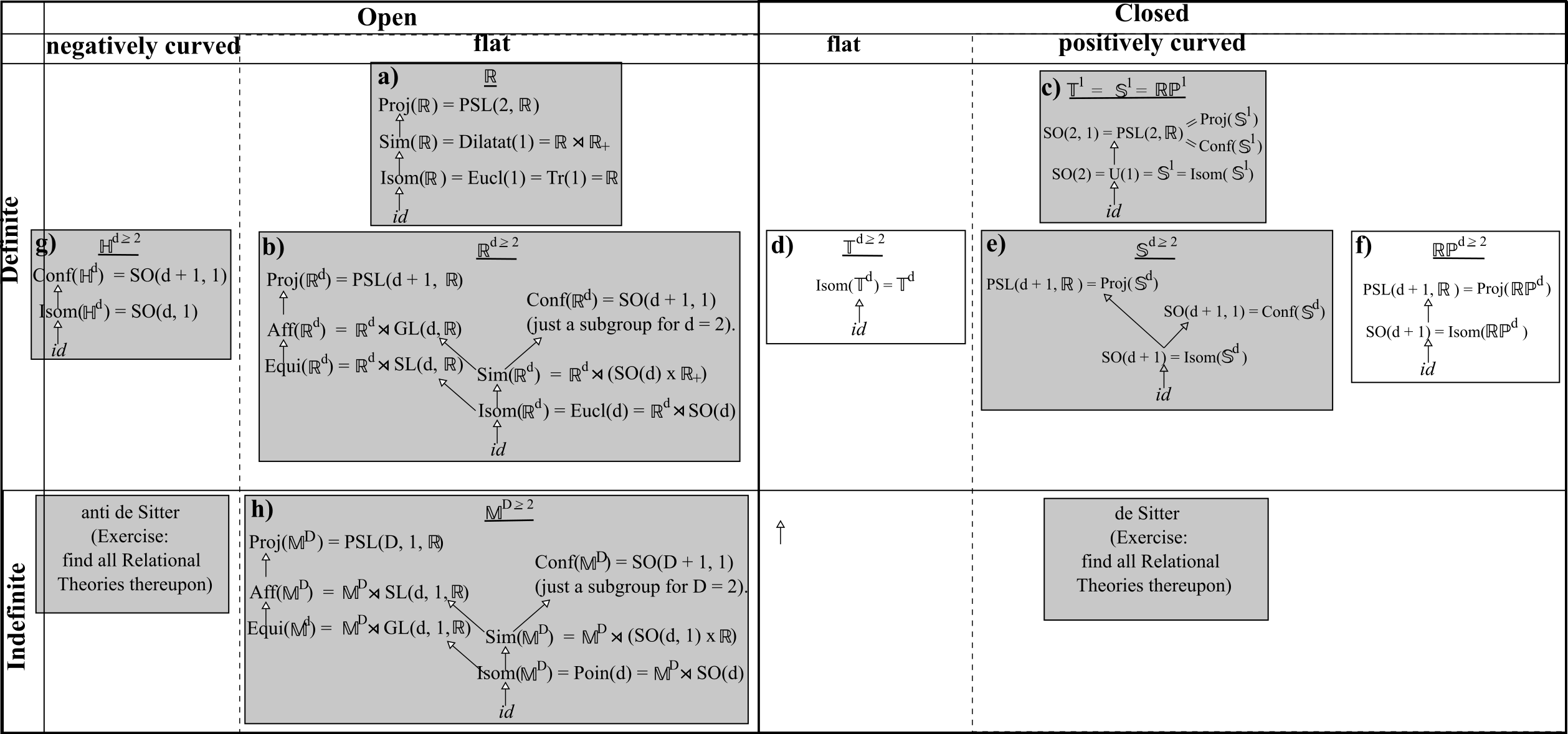}
\caption[Text der im Bilderverzeichnis auftaucht]{\footnotesize{Poset of geometrically significant automorphism groups yielding relational theories for 
a) $\mathbb{R}$, 
b) $\mathbb{R}^d$ for $d \geq 2$,
c) $\mathbb{T}^1 = \mathbb{S}^1 = \mathbb{RP}^1$, 
d) $\mathbb{T}^d$, 
e) $\mathbb{S}^d$, 
f) $\mathbb{RP}^d$ (each for $d \geq 2$),
g) $\mathbb{H}^d$, 
h) $\mathbb{M}^D$ for $D \geq 2$.   
The effect on diversity of nontrivial Relational Theories of introducing each of curvature, non-maximality and topological identification, is moreover indicated.  
Grey squares indicate which cases are maximally symmetric.
White squares, moreover, coincidentally, also involve topological identification. }} 
\label{Killing1}\end{figure} } 

\vspace{10in}

\subsection{Relational Theories are Nongeneric} 

\n The preceding subsection gives structural clarity, and moreover provides a systematic procedure for finding geometrically-significant automorphism groups.

\m 

\n We observe a rapid loss of $\mathbb{R}^n$'s complexities; any of moving away from maximal symmetry, 
                                                                                     introducing curvature, 
																				  or introducting topological identification 
rapidly cuts down on the number of geometrically-significant automorphism groups, and consequently of Relational Theories. 

\m

\n For instance, isometry groups for Sec \ref{O-C}'s Examples 6 to 10 are singletons rather than being partnered by distinct similarity groups.
Thus these examples' Shape-and-Scale Theories are singleton Relational Theories rather than having any corresponding Shape Theories. 
This is because the generator of dilations does not respect $\mathbb{T}^d$'s identification or $\mathbb{S}^d$ and $\mathbb{RP}^d$'s curvature scale.  
In fact, similarity Killing vectors that are not also Killing vectors are exceedingly rare \cite{Yano}.  
This ends suggestions that 'further quotienting out scale' might have some deeper significance.  

\m  

\n Let us next consider {\sl generic} geometrically-equipped differential manifolds.

\m 

\n{\bf Definition 1} A common sense of {\it manifold genericity} is 
\be
Isom(\FrM^d) = id                                                            \m .
\ee 
\n We now interpret this moreover more specifically as {\it metric-level genericity}.

\m

\n{\bf Remark 1} Metric genericity however does not suffice for our purpose of characterizing manifolds supporting no nontrivial relational theories. 
Our first generalization is as follows. 

\m 

\n{\bf Definition 2} {\it Geometric genericity}, 
\be 
\mbox{all } \m Gen\mbox{-}Isom(\FrC^d) = id                             \m . 
\ee
\n{\bf Remark 2} This generalization does not moreover cover how Projective Shape Theory is assigned. 
Including this case can require that 
\be 
Proj(\FrP^{d - 1}) = id \m  
\ee 
as well, for $\FrP^{d - 1}$ the {\it projectivized version} of the incipient carrier space. 
How far this definition extends is an issue that the continuing conceptual development of Projective Shape Theory will need to face; 
it is clear for flat space (including indefinite and Grassmannian cases). 

\m 

\n{\bf Definition 3} So we have a {\it genericity condition} consisting of {\it geometric genericity of carrier space} $\FrC^d$ 
                                                                           {\it and its projectivization} $\FrP^{d - 1}$.  

\m 
 
\n{\bf Proposition 1} For $\FrC^d$ generic, and using $\FrR$ to denote relational space with no implications made as to whether it is a shape or shape-and-scale space, 
\be 
\FrR(\FrC^d, N; Gen\mbox{-}Isom(\FrC^d))  \es  \frac{\FrQ(\FrC^d, N)}{Gen\mbox{-}Isom(\FrC^d)} 
                                          \es  \frac{\bigtimes_{I = 1}^{N}  \FrC^d}{id}
							              \es  \bigtimes_{I = 1}^{N}        \FrC^d                                   \m , 
\ee 
and (if defineable) 
\be 
\FrR(\FrP^{d - 1}, N; Gen\mbox{-}Isom(\FrP^{d - 1}))  \es  \frac{\FrQ(\FrP^{d - 1}, N)}{Gen\mbox{-}Isom(\FrP^{d - 1})} 
                               \es  \frac{\bigtimes_{I = 1}^{N}  \FrP^{d - 1}}{id}
							   \es  \bigtimes_{I = 1}^{N}        \FrP^{d - 1}                                          
\ee 
are the only options available for relational spaces, and corresponding relational theories. 
I.e.\ just the constellation space and the projectivize constellation space. 
{\sl Generically, distinct Shape Theories and Shape-and-Scale Theories are not supported}: 
an argument against ascribing fundamentality specifically to Shape Theories.  

\m 

\n{\bf Corollary 1} Both of the above relational spaces are moreover just {\sl product spaces}. 
{\sl Thus in the Mechanics case, quotienting by groups of continuous transformations is avoided altogether.} 
Indeed, such finite products of topological spaces preserve initial Hausdorffness, second-countability and local Euclideanness, 
so these constellation spaces inherit $\FrC^d$ and $\FrP^{d - 1}$'s manifoldness.

\section{Conclusion}

\n The diversity of Relational Theories -- Shape Theories and Shape-and-Scale Theories found to date is underlied by the poset of generalized Killing equations (GKEs) 
on each carrier space manifold. 
We take this into account by modifying the Introduction's structural outlay of Relational Theory as follows. 

\m 

\n We firstly insert {\bf Structure III.A}: the poset of GKEs on a given carrier space, 

\m 

\n Each of these equations is to be solved to obtain the corresponding geometrically-significant automorphism group, 
thus {\sl recovering} the Introduction's Structure III, albeit now relabelled {\bf Structure III.B}.

\m 

\n It is moreover well-known that these equations generically admit no nontrivial solutions on the geometrically-equipped differentiable manifolds. 
For these solutions are generalized Killing vectors, and such are only supported by geometrically-equipped differentiable manifolds in possession of generalized symmetries, 
of which the generic such manifolds have none.  

\m 

\n Thus, there are generically {\sl no} nontrivial Relational Theories 
(allowing also for its projectivised counterpart, where defineable, also possessing no gneralized symmetries).  
There is always a trivial Relational Theory: that whose relational space is just the constellation space product rather than a nontrivial quotients thereof. 
These correspond to the identity automorphism forming the trivial group; this much is always present since GKEs are homogeneous-linear and thus are solved by zero.  
This is a useful observation given recent diversification posited and at least partly solvable models in this subject 
\cite{Kendall, Sparr, MP03, MP05, FileR, Quad-I, Bhatta, AMech, PE16, I-II-III, ACirc, ASphe, ATorus, Aff}.  
Nontrivial Relational Theories are thus in fact very rare -- measure zero in the space of all $\langle \FrM, \bsigma \rangle$. 
The recent rapid expansion in such theories derived from searching in the most highly symmetric settings.
Their diversity moreover very rapidly falls off upon introduction of even just some of curvature (Differential Geometry), 
                                                                                       non-maximality of symmetry (Group Theory), 
																					or topological identification (Topology). 
This is the first great bound on the extent of Relational Theory, in the form of a Differential-Geometric level result, the GKE being rooted in the Lie derivative.  

\m 

\n{\bf Acknowledgments} I thank Malcolm MacCallum for teaching me about the importance of Killing vectors, and Chris Isham and Don Page for further discussions.  
I thank Don, Malcolm, Jeremy Butterfield, Enrique Alvarez and Reza Tavakol for support with my career. 
I finally dedicate this Article to my strongest friend.


\end{document}